\documentclass[nofootinbib,aps,preprint,superscriptaddress]{revtex4}
\usepackage{epsfig}
\usepackage{amsmath}

\newcommand{\fms}[1]{{#1}\!\!\!/}
\newcommand{\fmsl}[1]{{#1}\!\!\!\!/}
\newcommand{\n}{\overline{n}}

\newcommand{\mO}{\mathcal{O}}
\newcommand{\mP}{\mathcal{P}}

\newcommand{\mC}{\mathcal{C}}
\newcommand{\mB}{\mathcal{B}}
\newcommand{\mL}{\mathcal{L}}
\newcommand{\mT}{\mathcal{T}}
\newcommand{\mA}{\mathcal{A}}
\newcommand{\nn}{\frac{\fms{\overline{n}}}{2}} 
\newcommand{\nnn}{\frac{\fms{n}}{2}} 
\newcommand{\veps}{\varepsilon}
\newcommand{\eps}{\epsilon}

\newcommand{\SI}{\rm{SCET_I}}
\newcommand{\SII}{\rm{SCET_{II}}}

\begin{document}

\baselineskip 3.0ex 

\vspace*{18pt}

\title{Nonperturbative Charming Penguin Contributions to Isospin Asymmetries in Radiative $B$ decays}

\def\addDuke{Department of Physics, Duke University, Durham NC 27708, USA} 
\def\addPitt{Department of Physics and Astronomy, University of
  Pittsburgh, PA 15260, USA} 

\author{Chul Kim}\email{chul@phy.duke.edu}\affiliation{\addDuke}
\author{Adam K. Leibovich}\email{akl2@pitt.edu}\affiliation{\addPitt} 
\author{Thomas Mehen}\email{mehen@phy.duke.edu}\affiliation{\addDuke}

\begin{abstract} \vspace*{18pt}
\baselineskip 3.0ex 

Recent experimental data on the radiative decays $B \to V \gamma$, where $V$ is a light vector meson,
find small isospin violation in $B \to K^* \gamma$ while isospin asymmetries in $B \to \rho \gamma$
are of order 20\%, with large uncertainties. Using Soft-Collinear Effective Theory, we 
calculate isospin asymmetries in these radiative $B$ decays up to $O(1/m_b)$, also including
$O(v\alpha_s)$ contributions from nonperturbative charming penguins (NPCP). 
In the absence of NPCP contributions, the theoretical predictions for the asymmetries are a few percent or less.
Including the NPCP can significantly increase the isospin asymmetries for both $B \to V \gamma$ modes.
We also consider the effect of the NPCP on the branching ratio and CP asymmetries in $B^\pm \to V^\pm \gamma$.

\end{abstract}

\maketitle 

The rare radiative $B$ decays, $B \to V\gamma$, where $V$ is a light vector meson, are important  in  heavy flavor physics
because the dominant processes are due to the flavor changing neutral current. Isospin asymmetries are interesting
observables for testing the Standard Model (SM) and investigating new physics in the flavor sector. The isospin asymmetries
for $B\to K^*\gamma$ and $B\to \rho\gamma$ are defined to be
\begin{eqnarray} 
\Delta_{0-}^{K^*} &=& \frac{\Gamma(\overline{B}^0 \to \overline{K}^{*0} \gamma) - \Gamma(B^- \to K^{*-} \gamma)}{\Gamma(\overline{B}^0 \to \overline{K}^{*0} \gamma) + \Gamma(B^- \to K^{*-} \gamma)}, \nonumber \\  
\label{isoasym} 
\Delta_{0-}^{\rho} &=& \frac{2\Gamma(\overline{B}^0 \to \rho^{0} \gamma) - \Gamma(B^- \to \rho^{-} \gamma)}{2\Gamma(\overline{B}^0 \to \rho^{0} \gamma) + \Gamma(B^- \to \rho^{-} \gamma)}.
\end{eqnarray} 
In these asymmetries, the decay rates are averaged over charge conjugate modes.
Recent experimental measurements find \cite{Experiment}  
\begin{equation} 
\label{expasym}
\Delta_{0-}^{K^*} = 0.03\pm 0.04,~~~\Delta_{0-}^{\rho} = 0.26\pm 0.14, 
\end{equation} 
where the average values for the decay rates for $B \to K^*(\rho) \gamma$ are taken from the Heavy Flavor Averaging Group~\cite{HFAG}.  
The isospin asymmetry for $B\to K^*\gamma$ is consistent with zero within an error of a few percent.
The data suggests that the asymmetry in $B\to \rho \gamma$ is significantly larger, but 
because of large uncertainties   it is not yet possible to draw a definitive
conclusion. The work in this paper is motivated by the question of whether an anomalously large  isospin
asymmetry in $B\to \rho\gamma$ can be understood within the SM. In particular, we calculate
subleading contributions to the leading QCD factorization theorems for $B \to V \gamma$ to see if they can explain
the observed asymmetries.

\begin{figure}[b]
\begin{center}
\epsfig{file=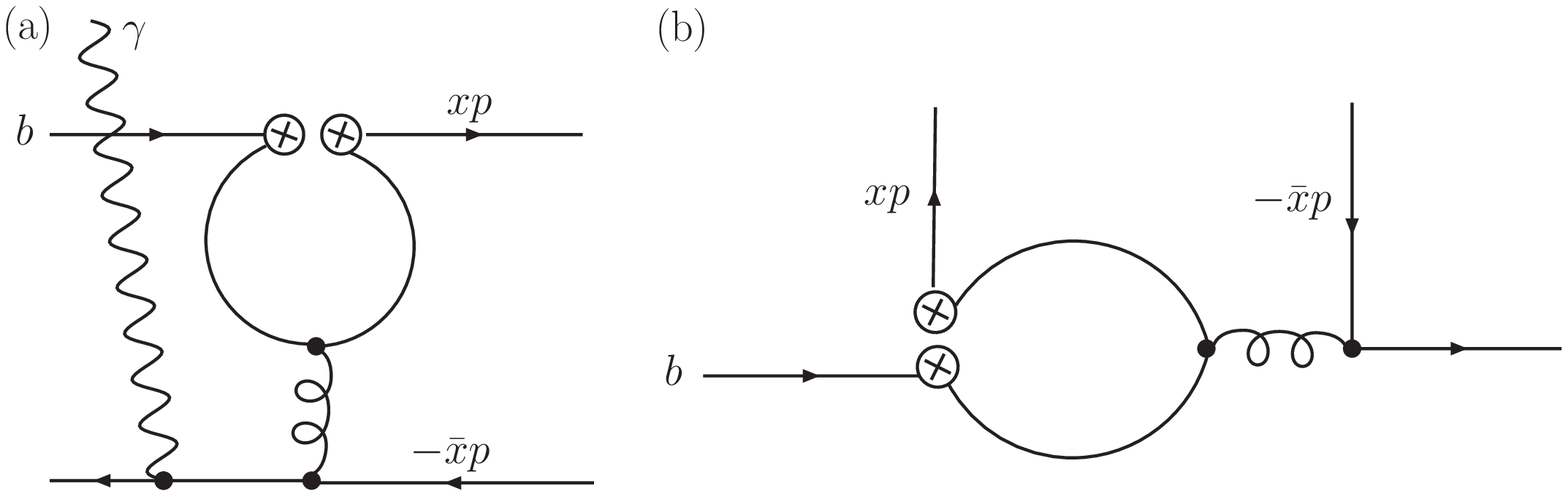, width=15cm, height=4.5cm}
\end{center}
\vspace{-0.7cm}
\caption{\baselineskip 3.0ex  
 Nonperturbative charming penguin (NPCP) contributions for (a) $B\to V \gamma$ and (b) $B\to M_1 M_2$ arise 
when $\bar{x}=1-x \approx 4 m_c^2/m_b^2$, in which case the charm quark pair is in the threshold region.
} 
\label{charmcomp} 
\end{figure}

In the heavy quark limit, the leading amplitudes are factorizable
\cite{Bosch:2001gv,Ali:2001ez,Chay:2003kb,Becher:2005fg,Ali:2007sj}.  However, isospin violating asymmetries come from
$O(1/m_b)$ suppressed power corrections for which  the factorization is necessarily more complicated. In this paper we
calculate $O(1/m_b)$ corrections to the asymmetries using Soft-Collinear Effective Theory (SCET) \cite{SCET}, which provides
a systematic power counting. In addition, possible endpoint divergences in these higher order corrections can be regulated
without imposing an arbitrary infrared cutoff by including the zero-bin subtraction of Ref.~\cite{Manohar:2006nz}.  Previous
QCD analyses of isospin asymmetries in radiative $B$ decays appear in Refs.~\cite{Kagan:2001zk,Beneke:2004dp,Donoghue:1996fv}. The main
difference between our analysis and previous work is the inclusion of nonperturbative charming penguin (NPCP) contributions,
which are already  known to play an important role in nonleptonic $B$ decays~\cite{Ciuchini:1997hb,Ciuchini:2001gv,Bauer:2005wb}. 
(For an alternative point of view, see Ref.~\cite{Beneke:2004bn}).

The NPCP contributions to $B \to V \gamma$ are depicted in Fig.~\ref{charmcomp}-(a).  In certain kinematic regimes, the invariant mass of the charm
quark pair  in the loop in Fig.~\ref{charmcomp}-(a) is near the threshold, $2 m_c$, in which case the charm quark pair is described by
nonrelativistic QCD (NRQCD) and additional interactions need to be taken into account. As pointed out in Ref.~\cite{Bauer:2005wb}, contributions 
from this regime are suppressed by only $v \alpha_s (2m_c)$  compared to the leading contribution. Here $v$ is the relative velocity 
of the charm quarks in the threshold region. Therefore, the NPCP contribution to the isospin
asymmetry could dominate over other $1/m_b$  suppressed contributions.  In this paper, we calculate the isospin 
asymmetries including the NPCP along 
with $1/m_b$ suppressed contributions.  We also calculate the NPCP contributions to 
the branching ratio and CP asymmetries for $B^\pm \to V^\pm \gamma$.

In the absence of NPCP contributions, the theoretical predictions for  $\Delta^{K^*}_{0-}$ and $\Delta^\rho_{0-}$
are no larger than a few percent. Including the NPCP contributions can significantly modify the theoretical predictions for the isospin
asymmetries. We will see below that the NPCP contribution can be factorized using SCET, and the result
expressed in terms of nonperturbative matrix elements.
The NPCP matrix elements are fitted to available data on the  isospin 
asymmetries, $\Delta^{K^*}_{0-}$ and $\Delta^\rho_{0-}$, the CP asymmetry for $B^\pm \to \rho^\pm \gamma$ 
(recently measured by Belle~\cite{Taniguchi:2008ty}), and the branching ratio for $B^+ \to \rho^+ \gamma$~\cite{HFAG}.
The predictions for  $\Delta^{K^*}_{0-}$ and $\Delta^\rho_{0-}$ are of order 10\%, 
with uncertainties large enough that both predictions are consistent with experiment.
However, the NPCP does not predict a large difference between $\Delta^{K^*}_{0-}$ and $\Delta^\rho_{0-}$, as suggested
by the central values  in Eq.~(\ref{expasym}).

The isospin asymmetry in $B \to V \gamma$ can arise from either the mass difference of the spectator quark in the 
$B$ meson or the electric charge difference when
the spectator quark emits the photon in the final state. However the isospin asymmetry due to the mass difference is negligible because it is
$\mO((m_u-m_d)/\Lambda_{\rm QCD})$ and therefore of order 1\% or smaller. So the dominant piece comes from electromagnetic (EM) interactions with the spectator
quark.


In order to describe the isospin breaking corrections to $B\to V \gamma$, we need the following effective weak Hamiltonian
\begin{equation} 
H_{W} = \frac{G_F}{\sqrt{2}} 
\Biggl[ \sum_{p=u,c}  \lambda_p^{(q)} \Bigl(C_1 O_{1p} + C_2 O_{2p}
\Bigr) - \lambda_t^{(q)} \Bigl( \sum_{i=3}^{6} C_i O_i + C_{8g} O_{8g} +
C_{7\gamma} O_{7\gamma}\Bigr) \Biggr], 
\label{fullw}
\end{equation} 
where the operators are 
\begin{alignat}{2} 
O_{1p} &= (\overline{p}b )_{V-A}
(\overline{q}p)_{V-A},\qquad\qquad\qquad\quad&  
O_{2p} &= (\overline{p}_{\beta}b_{\alpha} )_{V-A} 
(\overline{q}_{\alpha}p_{\beta} )_{V-A},  \nonumber \\  
O_{3,5} &= (\overline{q}b )_{V-A} \sum_{q'=u,d,s,c,b} (\overline{q'}q' )_{V\mp A},   
& O_{4,6} &= (\overline{q}_{\beta}b_{\alpha} )_{V-A} \sum_{q'=u,d,s,c,b} 
(\overline{q'}_{\alpha}q'_{\beta} )_{V\mp A}, \label{weakhamiltonian} \\   
O_{7\gamma} &= - \frac{em_b}{8\pi^2} \overline{q} 
\sigma^{\mu\nu} F_{\mu\nu} (1+\gamma_5) b, & 
O_{8g} &= - \frac{gm_b}{8\pi^2} \overline{q} \sigma^{\mu\nu} 
 G^a_{\mu\nu}T^a (1+\gamma_5) b. \nonumber 
\end{alignat}
Here $q$ is the $d$ or $s$ quark, the CKM factor is $\lambda_p^{(q)}= V_{pb}V_{pq}^*$, and $V\pm A = \gamma^{\mu} (1\pm \gamma_5)$.

\begin{figure}[t]
\begin{center}
\epsfig{file=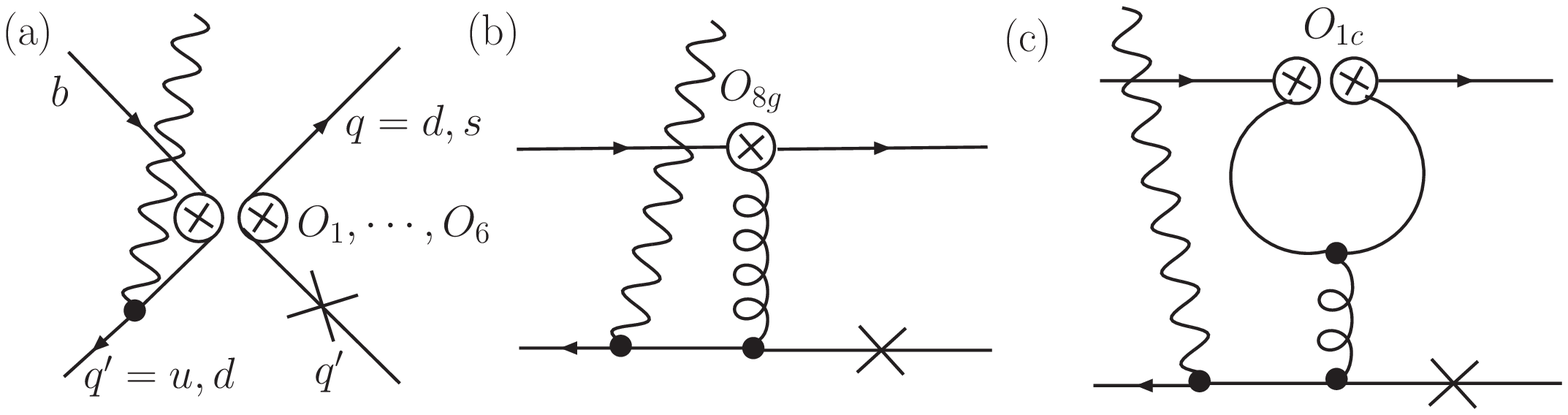, width=16cm, height=4cm}
\end{center}
\vspace{-0.7cm}
\caption{\baselineskip 3.0ex Various isospin breaking contributions in full QCD. Here crosses denote another possible photon emissions from the spectator quark. If we do not consider the long distance contribution in the diagram (c), all the contributions are power-suppressed by $\mO (\Lambda/m_b)$.   
 } 
\label{fullisoasym} 
\end{figure}

For the asymmetric contributions, the photon radiates from either the initial or final spectator quark as shown in Fig.~\ref{fullisoasym}.   If the
photon is radiated from the initial spectator quark, we need SCET operators of the type $\mO^{(0,4q)} = \overline{\chi}_{\n} \Gamma b_v
\overline{\xi}_n \Gamma \xi_n$  where $\xi_n(\chi_{\n})$ is $n(\n)$-collinear field\footnote{Our conventions are the same as Ref.~\cite{Chay:2006ve}
except that we have  exchanged  $n\leftrightarrow \n$ compared to that paper.} and $n$ and $\n$ are light-cone vectors satisfying $n^2=\n^2=0,~n\cdot
\n =2$. The analysis of factorization for these operators appears in Refs.~\cite{Chay:2003zp,Chay:2003ju} and the Wilson coefficients at
next-to-leading order (NLO) in $\alpha_s$ have been calculated in Refs.~\cite{Chay:2003ju,Chay:2006ve}. 

The leading operator in SCET only contributes to longitudinally polarized vector meson production, but in $B\to V \gamma$ the 
vector meson must be transversely polarized.   Transversely polarized vector mesons can be produced from subleading operators
that are higher order in the SCET expansion parameter, $\lambda$.  The relevant effective weak Hamiltonian in SCET is 
\begin{equation} 
\label{SCET4q} 
H_{W,\rm{SCET}}^{(1,4q)} = \frac{G_F}{\sqrt{2}} \sum_p \lambda_p^{(q)} \sum_{i=1}^6 \int^1_0 dx \mB_i^p (x,\mu) \mO^{(1,4q)}_i (x,\mu),
\end{equation} 
where $\mO^{(1,4q)}_i$ are  
\begin{eqnarray} 
\mO^{(1,4q)}_1 &=& \overline{\chi}_{\n}^u W_{\n} \gamma^{\perp}_{\mu} (1-\gamma_5) Y_{\n}^{\dagger} b_v  \Bigl[\overline{\xi}_{\n}^q W_n \gamma_{\perp}^{\mu} (1-\gamma_5) W_n^{\dagger} \xi_n^u + \overline{\xi}_{n}^q W_n \gamma_{\perp}^{\mu} (1-\gamma_5) W_n^{\dagger} \xi_{\n}^u\Bigr]_x, \nonumber \\
\label{o14qi} 
\mO^{(1,4q)}_{2,3} &=& \overline{\chi}_{\n}^q W_{\n} \gamma^{\perp}_{\mu} (1-\gamma_5) Y_{\n}^{\dagger} b_v  \Bigl[\overline{\xi}_{\n}^u W_n \gamma_{\perp}^{\mu} (1\mp\gamma_5) W_n^{\dagger} \xi_n^u + \overline{\xi}_{n}^u W_n \gamma_{\perp}^{\mu} (1\mp\gamma_5) W_n^{\dagger} \xi_{\n}^u\Bigr]_x,  \\
\mO^{(1,4q)}_4 &=& \sum_{q'=u,d,s} \overline{\chi}_{\n}^{q'} W_{\n} \gamma^{\perp}_{\mu} (1-\gamma_5) Y_{\n}^{\dagger} b_v  \Bigl[\overline{\xi}_{\n}^q W_n \gamma_{\perp}^{\mu} (1-\gamma_5) W_n^{\dagger} \xi_n^{q'} + \overline{\xi}_{n}^q W_n \gamma_{\perp}^{\mu} (1-\gamma_5) W_n^{\dagger} \xi_{\n}^{q'} \Bigr]_x, \nonumber \\
\mO^{(1,4q)}_{5,6} &=& \sum_{q'=u,d,s} \overline{\chi}_{\n}^{q} W_{\n} \gamma^{\perp}_{\mu} (1-\gamma_5) Y_{\n}^{\dagger} b_v  \Bigl[\overline{\xi}_{\n}^{q'} W_n \gamma_{\perp}^{\mu} (1\mp\gamma_5) W_n^{\dagger} \xi_n^{q'} + \overline{\xi}_{n}^{q'} W_n \gamma_{\perp}^{\mu} (1\mp\gamma_5) W_n^{\dagger} \xi_{\n}^{q'} \Bigr]_x \nonumber. 
\end{eqnarray}
Here the superscript `1' denotes suppression by one power of $\lambda$ compared to the leading operator,
$W_{n(\n)}$ is a collinear  Wilson line in $n(\n)$-direction, and $Y_{\n}$ is a ultrasoft (usoft) Wilson line. The
subscript outside the square brackets denotes that a delta function which fixes the momentum fraction, $x$, is included in the bilinear
operator:
\begin{equation} 
\Bigl[\overline{\xi}_{\bar n} W_n \Gamma W_n^{\dagger} \xi_n \Bigr]_x \equiv 
\overline{\xi}_{\bar n} W_n \delta\Bigl(x-\frac{\overline{\mP}^{\dagger}}{2E_V}\Bigr)\Gamma W_n^{\dagger} \xi_n, 
\end{equation} 
where $\overline{\mP} = \n\cdot \mP$ is a derivative operator taking the largest momentum component and $E_V$ is the energy of the produced vector
meson.  Here, $\xi_{\n}$ is the power-suppressed component in the spin projection of $q_n=(\fms{n}\fms{\n}/4) \xi_n + (\fms{\n}\fms{n}/4)  \xi_{\n}$, where
$q_n$ is the collinear quark field. Using the equation of the motion, $\xi_{\n}$ can be expressed in terms of $\xi_n$ as
\begin{equation} 
\xi_{\n} = W_n \frac{1}{\overline{\mP}} W_n^{\dagger} i\fmsl{D}_{\perp} \nn \xi_n.
\end{equation} 

\begin{figure}[b]
\begin{center}
\epsfig{file=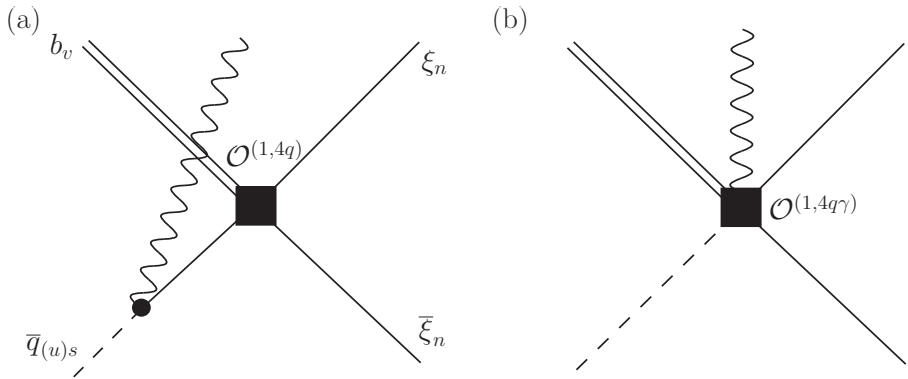, width=12cm, height=5cm}
\end{center}
\vspace{-0.7cm}
\caption{\baselineskip 3.0ex    
SCET diagrams for the isospin breaking corrections. Each diagram represent the electromagnetic interactions from initial and final spectator quarks, respectively. } 
\label{scetisoasym} 
\end{figure}

The Wilson coefficients $\mB_i^p$ in Eq.~(\ref{SCET4q}) are the same as the leading Wilson coefficients $\mC_i^p$ of Ref.~\cite{Chay:2006ve} due to 
reparameterization invariance \cite{RPI}. Because the isospin-breaking contributions from $H_{W,\rm{SCET}}^{(1,4q)}$ are suppressed by $\Lambda_{\rm QCD}/m_b$ compared to
the leading decay amplitude, we will suppress EM penguins with $C_{7,8,9,10}$ at tree level and keep only $C_{1,2,8g}$ at one loop order in $\mB_i^p$ for our
phenomenological analysis. 

The four-quark operators in the effective weak SCET Hamiltonian, Eq.~(\ref{SCET4q}), contribute to the radiative weak decay when the photon is emitted from the
initial spectator quark,  as shown in Fig.~\ref{scetisoasym}-(a).  To calculate their contribution, we need to take the time-ordered products of $\mO_i^{(1,4q)}$
with the electromagnetic interaction term $\mL^{(1)}_{\rm{EM}}$,
\begin{equation} 
\label{lem1} 
\mL^{(1)}_{\rm{EM}} = e_q \overline{q}_{us} Y_{\n} \fms{\rm{A}}_{\perp} W_{\n}^{\dagger} \chi_{\n}^q + \rm{h.c.}, 
\end{equation} 
where $\rm{A}^{\mu}$ is a photon field and $e_q$ is the electric charge of the quark. The time-ordered products are performed in $\SI$ with possible offshellness
$p^2 \sim m_b\Lambda$ and then matched onto $\SII$, which describes dynamics with fluctuations of $p^2 \sim \Lambda^2$. In the matching,  $\n$-collinear
fields having large offshellness of $\mO (m_b \Lambda)$ must  be integrated out, giving a jet function  
\begin{equation} 
\label{chijet} 
\langle 0 | \mathrm{T}\{W_{\n}^{\dagger} \chi_{\n} (z), \overline{\chi}_{\n} W_{\n} (0)\} | 0 \rangle = i \nn \delta(z_-) \delta^2 (z_{\perp}) \int \frac{dk_-}{2\pi} 
e^{-ik_-z_+/2} J_{n\cdot p_{\gamma}} (k_-), 
\end{equation}
where $k_+ = n\cdot k$, $k_- = \n\cdot k$, and the momentum of the photon,   $p_{\gamma}^{\mu} = n\cdot p_{\gamma} \n^{\mu}/2 = m_b \n^{\mu}/2$. At the lowest
order in $\alpha_s$, the jet function is simply $J_{n\cdot p_{\gamma}} = 1/k_-$. 

The $n$- and $\n$-collinear degrees of freedom are decoupled at leading order in SCET, and by using a field redefinition it is possible to decouple usoft degrees
of freedom from both.  Therefore the $n$-collinear piece of the matrix element describing the production of the light vector meson is decoupled from the
$\n$-collinear and usoft parts.  Thus, we can compute $B$ to $\gamma$ from $H_{W,\rm{SCET}}^{(1,4q)}$, which only depends on usoft and $\n$-collinear physics,
independently of the light-meson production process, which depends on $n$-collinear physics.  After a brief calculation, we find 
\begin{eqnarray} 
\label{T4q} 
\hat{\mT}_{4q}^{\mu} &=& i\int d^4 z \langle \gamma (\eps_{\perp}^*) | \mathrm{T}\{\overline{\chi}_{\n}^q W_{\n} \gamma^{\perp}_{\mu} (1-\gamma_5) Y_{\n}^{\dagger} b_v (0), \mL_{\rm{EM}}^{(1)} (z)\}| B \rangle, \\ 
&=& -i\frac{e_q}{2} f_B m_B (\eps_{\perp}^{*\mu} + i \veps_{\perp}^{\mu\nu} \eps_{\perp\nu}^*) \int dl_- J_{n\cdot p_{\gamma}} (-l_-) \phi_B^+ (l_-), \nonumber 
\end{eqnarray} 
where $\veps_{\perp}^{\mu\nu} = \veps^{\mu\nu\rho\sigma} n_{\rho} \n_{\sigma}/2$ setting $\veps^{0123} = -1$. $\phi_B^+$ is a light-cone distribution amplitude
(LCDA) of $B$ meson \cite{Grozin:1996pq}. We use the convention of Ref.~\cite{Beneke:2000wa} with $n$ and $\n$ interchanged.  

The light meson production is described by the $n$-collinear part of the matrix elements.  The matrix elements for production of transversely polarized vector
mesons in SCET are 
\begin{eqnarray} 
\label{gpv} 
\langle V_{\perp} (\eta_{\perp}^*) | \Bigl[\overline{\xi}_{\n} W_n \gamma_{\perp}^{\mu}  W_n^{\dagger} \xi_n + \overline{\xi}_{n} W_n \gamma_{\perp}^{\mu} W_n^{\dagger} \xi_{\n}\Bigr]_x |0 \rangle &=& -if_V m_V \eta_{\perp}^{*\mu} g_{\perp}^{(v)} (x), \\ 
\label{gpa}
\langle V_{\perp} (\eta_{\perp}^*) | \Bigl[\overline{\xi}_{\n} W_n \gamma_{\perp}^{\mu} \gamma_5 W_n^{\dagger} \xi_n + \overline{\xi}_{n} W_n \gamma_{\perp}^{\mu} \gamma_5 W_n^{\dagger} \xi_{\n}\Bigr]_x |0 \rangle &=& - \frac{f_V}{4} m_V \veps_{\perp}^{\mu\nu} \eta_{\perp\nu}^{*} \frac{\partial}{\partial x} g_{\perp}^{(a)} (x), 
\end{eqnarray}
where $g_{\perp}^{(v,a)}$ are chiral-even, twist-3 LCDAs \cite{Ball:1998sk} whose asymptotic forms are $g_{\perp}^{(v)} (x) = 3(x^2 + \overline{x}^2)/2$ and
$g_{\perp}^{(a)} (x) = 6x\overline{x}$, where $\overline{x} =1-x$. 

Combining Eqs.~(\ref{T4q}), (\ref{gpv}), and (\ref{gpa}),  the matrix element of $\mO_1^{(1,4q)}$ for $B^-\to\rho^-\gamma$, for example, is
\begin{eqnarray}  
\langle \rho^- \gamma | \mO_1^{(1,4q)} (x,\mu) | B^- \rangle &=& 
-\frac{e_u}{2} f_B f_{\rho} m_B m_{\rho} A_L (\eps_{\perp}^*, \eta^*_{\perp}) \int dl_- J_{n\cdot p_{\gamma}} (-l_-,\mu,\mu_0) \phi_B^+ (l_-,\mu_0) \nonumber \\ 
\label{mo4q}
&&\times \Bigl[g_{\perp}^{(v)} (x,\mu) - \frac{\partial}{4\partial x} g_{\perp}^{(a)} (x,\mu)\Bigr],
\end{eqnarray} 
where the renormalization scales are roughly given by $\mu \sim \sqrt{m_b\Lambda_{\rm QCD}}$ and $\mu_0 \sim
\Lambda_{\rm QCD}$, and $A_L = \eps_{\perp}^*\cdot \eta_{\perp}^* - i\veps_{\perp}^{\mu\nu}
\eps_{\perp\mu}^*\eta_{\perp\nu}^*$ is the  polarization factor for the left-handed $\rho^-$ and $\gamma$. 
In the case of $B\to \rho \gamma$, it is necessary to include one-loop corrections to the 
jet function, see Ref.~\cite{Lunghi:2002ju} for details.

When the photon radiates from a final state quark (i.e., from the crosses in Fig.~\ref{fullisoasym}), the
intermediate quark line is hard with offshellness  of order $m_b^2$.  Matching onto SCET we obtain localized
four-quark operators with photons, shown in Fig.~\ref{scetisoasym}-(b). In this case, the  effective
weak Hamiltonian that contributes to the decay amplitudes is 
\begin{equation} 
\label{SCET4qg}
H_{W,\rm{SCET}}^{(1,4q\gamma)} = \frac{G_F}{\sqrt{2}} \sum_p \lambda_p^{(q)} \sum_{i=1}^2 \int^1_0 dx \mA_i^p (x,\mu) \mO^{(1,4q\gamma)}_i (x,\mu),
\end{equation} 
where the five-particle operators $\mO^{(1,4q\gamma)}_i$ are 
\begin{equation} 
\label{o14qg}
\mO^{(1,4q\gamma)}_{\{1,2\}} (x) = \sum_{q'=u,d,s} e_{q'} \overline{q'}Y_n \{ \fms{\n},\fms{n}\} (1+\gamma_5) Y_n^{\dagger} b_v \Bigl[\overline{\xi}_n^q W_n \nn \fms{\mathrm{A}}_{\perp} (1+\gamma_5) \frac{1}{\overline{\mP}} W_n^{\dagger} \xi_n^{q'}\Bigr]_x.  
\end{equation} 
In Eq.~(\ref{SCET4qg}), the Wilson coefficients $\mA_i^p$ at NLO are
\begin{eqnarray} 
\label{Aip}
\mA_1^p (x,\mu) &=& C_6 + \frac{C_5}{N} + \frac{\alpha_s}{4\pi}\frac{C_F}{N}\Biggl\{\frac{2C_1}{3}\Bigl[1+\ln\frac{m_B^2}{\mu^2}-\frac{3}{2}G(s_p,x)\Bigr]-2C_{8g}\frac{m_b}{\bar{x}m_B} \Biggr\}, \\
\mA_2^p (x,\mu) &=& C_6 + \frac{C_5}{N} + \frac{\alpha_s}{4\pi}\frac{C_F}{N}\Biggl\{\frac{C_1}{3}\Bigl[1+\ln\frac{m_B^2}{\mu^2}-\frac{3}{2}G(s_p,x)\Bigr]-C_{8g}\frac{m_b}{m_B} \Biggr\},\nonumber 
\end{eqnarray} 
where $G(s_p,x)$  is
\begin{equation} 
G(s_p,x) = -4 \int^1_0 dz z\bar{z} \ln(s_p-z\bar{z}\bar{x}-i\eps),
\end{equation} 
and $s_p\equiv m_p^2/m_B^2$.
Similar to $\mB_i^p$, we neglect EM penguins at tree level and only keep $C_{1,2,8g}$ at one loop. The sum of the terms in Eq.~(\ref{Aip}) proportional to $C_1$
differs by a factor of 3/4 from Ref.~\cite{Kagan:2001zk}.

Again, the $B\to\gamma$ piece of the matrix element factors from the light-meson production piece.  The $n$-collinear part in Eq.~(\ref{o14qg}), describing the
vector meson production, gives a leading twist LCDA, $\phi_{\perp} (x)$, whose asymptotic form is $6x\bar{x}$. It can be obtained from the following projection:
\begin{equation} 
\label{phiperp} 
\Bigl\langle V_{\perp} (\eta_{\perp}^*) \Bigl| \Bigl[\overline{\xi}_n W_n \delta\Bigl(x-\frac{\overline{\mP}^{\dagger}}{2E_V}\Bigr) \Bigr]_a^{\alpha} \Bigl[W_n^{\dagger} \xi_n\Bigr]_b^{\beta} \Bigr| 0 \Bigr\rangle_{twist-2} = -\frac{i}{2} f_V^{\perp} E_V \Bigl(\fms{\eta}_{\perp}^* \nnn \Bigr)_{ba} \frac{\delta^{\beta\alpha}}{N} \phi_{\perp} (x).
\end{equation} 
As an example, the $B\to K^*\gamma$ matrix elements from $\mO^{(1,4q\gamma)}_i$ are 
\begin{equation} 
\label{mO14qg} 
\langle \overline{K}^{*0} \gamma | \mO^{(1,4q\gamma)}_1  | \overline{B}^0 \rangle = \langle \overline{K}^{*0} \gamma | \mO^{(1,4q\gamma)}_2  | \overline{B}^0 \rangle
= -\frac{e_d}{2} f_B f_{K^*}^{\perp} m_B A_L (\eps_{\perp}^*, \eta_{\perp}^*) \frac{\phi_{\perp} (x,\mu)}{\bar{x}}. 
\end{equation} 

Next we turn to the contributions  from  NPCP, shown in Fig.~\ref{charmcomp}-(a). 
The size of this contribution is $\mO(v\alpha_s(2m_c))$~\cite{Bauer:2005wb} and therefore is suppressed only logarithmically in the 
large $m_c$ limit, compared to the power suppression of previously considered $\mO (1/m_b)$ contributions. Numerically, $\Lambda_{\rm QCD}/m_b$
 and $v \alpha_s(2 m_c)$ are roughly the same size, so {\it a priori} it is sensible to include them at the same order. In fact we will see 
 below that the NPCP  gives the dominant contribution to isospin violation in $B \to \rho \gamma$.
  When $\bar{x}$ is close to $4s_c^2$, long-distance interactions govern the charm quark pair in the loop and hence it cannot be
separated from the $B$ meson.  However, the $n$-collinear piece of $V_{\perp}$ can still be decoupled, with the dominant part  obtained from the leading twist
projection of Eq.~(\ref{phiperp}). The factorization process is similar to the treatment in Ref.~\cite{Chay:2006ve} 
and we refer the reader to that paper for
details. The NPCP contribution to the decay amplitude is 
\begin{eqnarray} 
\label{HWcbc}
M^{c\bar{c}} &=& \frac{G_F}{\sqrt{2}} \lambda_c^{(q)}\langle V_{\perp} \gamma |  C_1 O_{1c} | B \rangle_{\rm{NPCP}} \\
&=& i\frac{G_F}{4N\sqrt{2}} \lambda_c^{(q)} f_{V}^{\perp} m_B \eta_{\perp\mu}^* \int dx \delta(\bar{x} -4s_c^2) \phi_{\perp} (x) 
H_{c\bar{c}} (x,m_B) \langle \gamma |\mO_{c\bar{c}}^{\mu} | B \rangle, \nonumber 
\end{eqnarray} 
where $H_{c\bar{c}} = 4C_1 \pi\alpha_s /(\bar{x} m_B^2)$ at lowest order and the six-quark operator 
$\mO_{c\bar{c}}^{\mu}$, including nonrelativistic charm quark fields, $c_{\pm\rm{v}}$, is defined as 
\begin{eqnarray} 
\mO_{c\bar{c}}^{\mu} &=& i\int d^4 y ~\overline{c}_{-\rm{v}} Y_n \gamma_{\perp}^{\nu} T^a Y_{\n}^{\dagger} c_{+\rm{v}} (y)  ~\overline{\chi}_{\n}^{q'} W_{\n} (y) \gamma^{\perp}_{\nu} \gamma^{\mu}_{\perp} \nnn \gamma^{\rho} (1-\gamma_5) T^a Y_n^{\dagger} c_{-\rm{v}} \nonumber \\
\label{Ocbc}
&&\times \overline{c}_{+\rm{v}} \gamma_{\rho} (1-\gamma_5) b_v (0).
\end{eqnarray}
In order to integrate out $\n$-collinear fields with fluctuations greater than $\Lambda_{\rm QCD}$, we consider time-ordered products of $\mO_{c\bar{c}}^{\mu}$
 with $\mL^{(1)}_{\rm{EM}}$,  
\begin{eqnarray}  
\label{Tcbc}
\hat{\mT}_{c\bar{c}} &=& i\eta_{\perp\mu}^* \int d^4 z \langle \gamma (\eps_{\perp}^*) |\mathrm{T}  \{\mO_{c\bar{c}}^{\mu} (0), \mL^{(1)}_{\rm{EM}} (z) \} | B \rangle\\
&=&ie_{q'} \int d^4 y \frac{dz_+dk_-}{4\pi} e^{-ik_-(z_+-y_+)/2} J_{n\cdot p_{\gamma}} (k_-) \langle 0 |\mO_{c\bar{c}q'b} (\eps_{\perp}^*, \eta_{\perp}^*,z_+,y) | B \rangle, \nonumber 
\end{eqnarray} 
where we employed Eq.~(\ref{chijet}) to obtain the second line of Eq.~(\ref{Tcbc}) and $\mO_{c\bar{c}q'b}$ is\begin{eqnarray} 
\label{Ocbcqb}
\mO_{c\bar{c}q'b} (\eps_{\perp}^*, \eta_{\perp}^*,z_+,y) &=& \overline{q'}_{us} Y_{\n} (\frac{z_+}{2},y_{\perp},\frac{y_-}{2}) \nn \fms{\eps}_{\perp}^*  \gamma_{\perp}^{\nu} \fms{\eta}_{\perp}^* \nnn  \gamma^{\rho} (1-\gamma_5) T^a Y_n c_{-\rm{v}} (0) \\
&&\times \overline{c}_{-\rm{v}} Y_n \gamma_{\nu}^{\perp} T^a Y_{\n}^{\dagger} c_{+\rm{v}} (y) 
\overline{c}_{+\rm{v}} \gamma_{\rho} (1-\gamma_5) b_v (0). \nonumber 
\end{eqnarray} 

The matrix element of $\mO_{c\bar{c}q'b}$ in Eq.~(\ref{Tcbc}) is purely nonperturbative. It can be 
decomposed into left- and right-handed polarized contributions,
\begin{eqnarray} 
\label{Fcbc}
&&\int d^4y \langle 0 | \mO_{c\bar{c}q'b} (\eps_{\perp}^*, \eta_{\perp}^*,z_+,y) | B \rangle e^{ik_- y_+/2} = f_B m_B^3 \int dl_- e^{-il_-z_+/2} \\
&&~~~~~~~~\times [ A_L (\eps_{\perp}^*,\eta_{\perp}^*) F_{c\bar{c}}^L (k_-,l_-)   
+A_R (\eps_{\perp}^*,\eta_{\perp}^*) F_{c\bar{c}}^R (k_-,l_-)], \nonumber 
\end{eqnarray} 
where $A_R = \eps_{\perp}^* \cdot \eta_{\perp}^* + i\veps_{\perp}^{\mu\nu} \eps_{\perp\mu}^* \eta_{\perp\nu}^*$  is the polarization factor for decay into
right-handed final states. An important point is that the NPCP can give a  right-handed polarized contribution which is not $\mO (1/m_b)$ suppressed.  
As pointed out in Ref.~\cite{Grinstein:2004uu}, any other right-handed polarized contributions to the decay amplitude should be suppressed by $1/m_b$. 
Therefore NPCP could give the dominant contribution to the right-handed polarized decay amplitudes.  
Since  the right-handed polarized contribution from NPCP cannot interfere with the leading order amplitude which produces only left-handed final states, the right-handed contribution does not enter into the asymmetries until higher orders.   Therefore, we neglect any possible
right-handed contribution from NPCP in our calculations of the asymmetries.

Combining Eqs.(\ref{HWcbc}), (\ref{Tcbc}), and (\ref{Fcbc}), we obtain 
\begin{equation}
\label{HWcbcf} 
M^{c\bar{c}} = -\frac{G_F}{\sqrt{2}} \lambda_c^{(q)} e_{q'} f_B f_V^{\perp} m_B^2 A_L (\eps_{\perp}^*,\eta_{\perp}^*)  \frac{\pi\alpha_s}{N\Lambda_{c\bar{c}}} \int^1_0 dx \frac{\phi_{\perp} (x)}{\bar{x}} \delta (\bar{x} -4s_c^2) \hat{H}_{c\bar{c}} (\bar{x}),
\end{equation} 
where $q=d$ or $s$, $q'$ is the $B$ meson spectator quark, and $\hat{H}_{c\bar{c}} = \bar{x} m_B^2
H_{c\bar{c}}/(4\pi \alpha_s) = C_1 + \cdots $.  Here we have defined  $\Lambda_{c\bar{c}}^{-1}$ to be
\begin{eqnarray} 
\label{Lcbc} 
\Lambda_{c\bar{c}}^{-1} &=& -\int dl_- \frac{dz_+dk_-}{4\pi} e^{-i(k_-+l_-)z_+/2} J_{n\cdot p_{\gamma}} (k_-) F_{c\bar{c}}^L (k_-,l_-) \\
&=&  -\int dl_-  J_{n\cdot p_{\gamma}} (-l_-) F_{c\bar{c}}^L (-l_-,l_-) \sim  \int dl_- \frac{F_{c\bar{c}}^L (-l_-,l_-)}{l_-}. \nonumber 
\end{eqnarray} 
Following Ref.~\cite{Bauer:2005wb}, we can power count the size of this correction.  
The NPCP contribution is suppressed relative to the leading order term by $v\alpha_s(2m_c)$, 
and thus of order $m_b v\alpha_s(2m_c)/\Lambda_{\rm QCD}$ relative to the other isospin breaking 
terms considered. Based on this power counting, we expect that 
\begin{equation}\label{npcpest}
\frac{m_B}{\Lambda_{c\bar{c}}} \frac{\phi_\perp(4 s_c^2)}{4s_c^2}
\sim \frac{v\,m_b}{\Lambda_{\rm QCD}} . 
\end{equation}
The factor $\phi_\perp(4 s_c^2)/(4s_c^2)$ is formally $O(1)$ in the power counting of Ref.~\cite{Bauer:2005wb},
but numerically $\phi_\perp(4 s_c^2)/(4s_c^2) \approx 4.3$, so we keep this factor in estimating $m_B/\Lambda_{c\bar{c}}$.
Taking $v\, m_b/\Lambda_{\rm QCD} \sim 3$ we find $m_B/\Lambda_{c\bar{c}} \sim 0.7$. The extracted values of $m_B/\Lambda_{c\bar{c}}$
are consistent with this naive estimate but smaller. For these values of $m_B/\Lambda_{c\bar{c}}$, the NPCP
gives significant contributions to the isospin asymmetries.

\begin{figure}[t]
\begin{center}
\epsfig{file=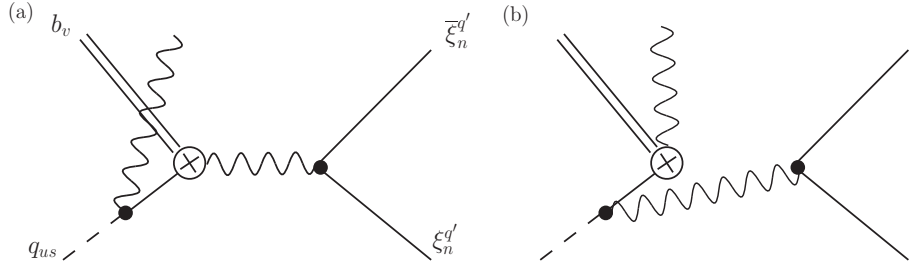, width=12cm, height=3.5cm}
\end{center}
\vspace{-0.7cm}
\caption{\baselineskip 3.0ex  
Leading double photon contribution to the isospin asymmetry, where $\otimes$ represents $O_{7\gamma}$. They are suppressed by $\alpha m_b^2/m_V^2$ compared to other usual isospin breaking contributions. } 
\label{doublephoton} 
\end{figure}

Finally, there is another interesting isospin-breaking source, a double photon contribution with the EM penguin
$O_{7\gamma}$. It is only available for the decay with an unflavored vector meson, i.e.,  $B \to \rho^0 \gamma$. The largest contributions are depicted in Fig.~\ref{doublephoton}. Concentrating first on Fig.~\ref{doublephoton}-(a), the off-shell photon coming from $O_{7\gamma}$ produces the vector meson and then an additional photon is emitted from the $B$ meson spectator quark.  Integrating out the hard photon, we can match onto the $\SI$ operator
\begin{eqnarray} 
\label{O2ga}
C_{7\gamma} O_{7\gamma} &\longrightarrow& \frac{ee_{q'}}{4\pi^2}\frac{m_bm_B}{m_V^2} \int dx \mC_{\gamma\gamma} (x) ~\overline{\chi}_{\n}^q W_{\n} \nnn \gamma_{\perp}^{\mu} (1+\gamma_5) Y_{\n}^{\dagger} h_v \\
&&\times \Bigl[\overline{\xi}_{\n}^{q'} W_n \gamma_{\mu}^{\perp} W_n^{\dagger} \xi_n^{q'} + \overline{\xi}_{n}^{q'} W_n \gamma_{\mu}^{\perp} W_n^{\dagger} \xi_{\n}^{q'} \Bigr]_x, \nonumber 
\end{eqnarray}  
where $C_{\gamma\gamma}$ is equal to $C_{7\gamma}$ at tree level. Next we integrate out the $\n$-collinear fields in the time-ordered product with $\mL_{\rm{EM}}^{(1)}$ and match onto $\SII$.  Applying Eqs.~(\ref{chijet}), (\ref{T4q}), and (\ref{gpv}),  we find 
\begin{equation} 
\label{mO2ga}
\langle V_{\perp} \gamma | C_{7\gamma} O_{7\gamma} | B \rangle_{2\gamma(\mathrm{a})}  = \frac{ee_qe_{q'}}{8\pi^2} f_B f_V \frac{m_b m_B^2}{m_V} C_{7\gamma} A_L (\eps_{\perp}^*,\eta_{\perp}^*) \int dl_- J_{n\cdot p_{\gamma}} (-l_-) \phi_B^+ (l_-), 
\end{equation} 
and, in case of $B \to \rho^0 \gamma$,
\begin{equation} 
\langle \rho^0_{\perp} \gamma | C_{7\gamma} O_{7\gamma} | \overline{B}^0 \rangle_{2\gamma(\mathrm{a})}
= \frac{Q_d(Q_u-Q_d)}{\sqrt{2}} \frac{e\alpha}{2\pi} f_B f_{\rho} \frac{m_b m_B^2}{m_V} C_{7\gamma} A_L (\eps_{\perp}^*,\eta_{\perp}^*) \int dl_- J_{n\cdot p_{\gamma}} (-l_-) \phi_B^+ (l_-), 
\end{equation} 
where $Q_q = e_q/e$ with $Q_u=2/3$ and $Q_d =-1/3$. 

At the lowest order in $\alpha_s$, the contribution of Fig.~\ref{doublephoton}-(b) is the same as Fig.~\ref{doublephoton}-(a) with $n$ and $\bar{n}$ exchanged. So the result can be written as 
\begin{equation} 
\langle \rho^0_{\perp} \gamma | C_{7\gamma} O_{7\gamma} | \overline{B}^0 \rangle_{2\gamma(\mathrm{b})}
= \frac{Q_d(Q_u-Q_d)}{\sqrt{2}} \frac{e\alpha}{2\pi} f_B f_{\rho} \frac{m_b m_B^2}{m_V} C_{7\gamma} A_L (\eps_{\perp}^*,\eta_{\perp}^*) \int dl_+ J_{n\cdot p_{\gamma}} (-l_+) \phi_B^+ (l_+). 
\end{equation}
The double-photon contribution is suppressed by $\alpha$, but enhanced by a factor of $m_b^2/m_V^2$ due to the virtual photon.
Compared to the other isospin-breaking breaking contributions, such as those in Eqs.~(\ref{mo4q}) and (\ref{mO14qg}), this contribution is rather
small. 

The isospin asymmetries in Eq.~(\ref{isoasym}) are given by
\begin{equation}\label{isas}
\Delta_{0-}^V = \frac{\mathrm{Re}(b_d^V - b_u^V) + R \, \mathrm{Re}(\bar{b}_d^V - \bar{b}_u^V)}{1+R},
\end{equation}
where
\begin{equation} 
\label{bi}
b_d^V= \frac{A_0^V}{c_V L_V},~~~b_u^V=\frac{A_-^V}{L_V}, 
\end{equation}
 $A_{0,-}^V$ are the leading isospin breaking corrections to the decay amplitude,
$L_V$ are the leading isospin symmetric decay amplitudes, and $c_V=1$ for $K^*$, $c_V= -1/\sqrt{2}$ for $\rho$.
In Eq.~(\ref{isas}),
 $\bar{b}^V_{u,d}$ are the corresponding ratios for the charge conjugate modes, and $R =|\overline{L_V}|^2/|L_V|^2$.
$L_V$ can be written as 
\begin{eqnarray} 
\label{leadingamp} 
L_{K^*} &=& \frac{G_F}{\sqrt{2}} \frac{e}{4\pi^2} m_b^2 m_B A_L(\eps_{\perp}^*, \eta_{\perp}^*) \lambda_c^{(s)} a_{7,K^*}^c \zeta_{\perp}^{K^*}, \\
L_{\rho} &=& \frac{G_F}{\sqrt{2}} \frac{e}{4\pi^2} m_b^2 m_B A_L(\eps_{\perp}^*, \eta_{\perp}^*) \sum_{p=u,c} \lambda_p^{(d)} a_{7,\rho}^p ~\zeta_{\perp}^{\rho}, \nonumber 
\end{eqnarray} 
where the transition form factor for $B \to V$, $\zeta_{\perp}^V$, is defined as 
\begin{equation} 
\langle V(\eta_{\perp}^*) | \overline{\xi}_n W_n \gamma_{\perp}^{\mu} (1-\gamma_5) Y_n^{\dagger} b_v | B\rangle = m_B (i\veps_{\perp}^{\mu\nu}\eta_{\perp\nu}^* - \eta_{\perp}^{*\mu}) \zeta_{\perp}^V,
\end{equation}  
and we will use $\zeta_{\perp}^{K^*} = 0.36\pm0.07$ and $\zeta_{\perp}^{\rho} = 0.27\pm0.05$ \cite{Beneke:2004dp} for the numerical analysis. 
The Wilson coefficients $a_{7,V}^p$ in Eq.~(\ref{leadingamp}) are  
\begin{eqnarray} 
\label{a7p}
a_{7,V}^p &=& C_{7\gamma} A_{7}^{(0)} +\frac{\alpha_s C_F}{4\pi} [C_1  G_1 (s_p) + C_{8g} G_8 (s_p) ] \\
&&+\pi\alpha_s \frac{C_F}{N} \frac{f_Bf_V^{\perp}m_B}{m_b^2} \int dl_+ \frac{\phi_B^+(l_+)}{l_+}\int^1_0 dx \frac{\phi_{\perp} (x)}{\bar{x}} \Bigl[C_{7\gamma} + \frac{C_1}{6} H(x,s_p) + \frac{C_{8g}}{3} \Bigr] \frac{1}{\zeta_{\perp}^V}, \nonumber
\end{eqnarray}
where the hard functions $A_7^{(0)}$, $G_{1,8}$, and $H$ are available in Refs.~\cite{Bosch:2001gv,Chay:2003kb,Greub:1996tg}, and we followed the conventions of
Ref.~\cite{Chay:2003kb}. 

Finally we obtain
\begin{eqnarray} 
\label{bqK}
&&b_q^{K^*} = Q_q \frac{2\pi^2 f_B}{m_b a_{7,K^*}^c \zeta_{\perp}^{K^*}}
\Bigl[2\frac{f_{\perp}^{K^*}}{m_b} K_1^{K^*} + \frac{f_{K^*} m_{K^*}}{\lambda_B m_b}K_{2q}^{K^*}\Bigr], \\
\label{bqrho}
&&b_q^{\rho} = Q_q \frac{2\pi^2 f_B}{m_b \sum_{p=u,c} \lambda_p^{(d)} a_{7,\rho}^p \zeta_{\perp}^{\rho}}
\Bigl[2\frac{f_{\perp}^{\rho}}{m_b} K_1^{\rho} + \frac{f_{\rho} m_{\rho}}{\lambda_B m_b} K_{2q}^{\rho}\Bigr].  
\end{eqnarray} 
Here $\lambda_B^{-1}=\int dl \phi_B^+(l)/l$ and we use the following model for $\phi_B^+(l)$  \cite{Braun:2003wx}
\begin{equation} 
\label{Bmodel} 
\phi_B^+ (l,\mu) =  \frac{4\mu l}{\pi\lambda_B(\l^2+\mu^2)}\Bigl[\frac{\mu^2}{l^2+\mu^2}
-\frac{2(\sigma_B-1)}{\pi^2} \ln\frac{l}{\mu}\Bigr], 
\end{equation} 
where the parameters $\lambda_B$ and $\sigma_B$ are $\lambda_B=460\pm110$ MeV, $\sigma_B =1.4\pm0.4$ at $\mu=1$ GeV. 
 $K_{1,2}$ can be written as 
\begin{eqnarray} 
\label{K1K} 
K_1^{K^*} &=& \int^1_0 dx \frac{\phi_{\perp} (x)}{\bar{x}} \Bigl\{-\frac{1}{2} \Bigl[\mA_1^c (x) + \mA_2^c (x)\Bigr] - C_1 \frac{\pi\alpha_s}{N} \frac{m_B}{\Lambda_{c\bar{c}}} \delta(\bar{x}-4s_c^2) \Bigr\} \\
\label{K2K} 
K_{2q}^{K^*} &=& \int^1_0 dx \Bigl[g_{\perp}^{(v)} -\frac{\partial}{4\partial x}  g_{\perp}^{(a)}\Bigr](x) 
\Bigl\{ \frac{\lambda_u^{(s)}}{\lambda_c^{(s)}}\Bigl(C_1+\frac{C_2}{N}\Bigr) \delta_{qu} + \mB_4^c (x,m_b)\Bigr\}, \\
\label{K1rho} 
K_1^{\rho} &=& \sum_{p=u,c} \lambda_p^{(d)} \int^1_0 dx \frac{\phi_{\perp} (x)}{\bar{x}} \Bigl\{-\frac{1}{2} \Bigl[\mA_1^p (x) + \mA_2^p (x)\Bigr] - \delta_{pc} C_1 \frac{\pi\alpha_s}{N} \frac{m_B}{\Lambda_{c\bar{c}}} \delta(\bar{x}-4s_c^2) \Bigr\}, \\
\label{K2rho}
K_{2q}^{\rho} &=& \sum_{p=u,c} \lambda_p^{(d)} \Biggl\{\int^1_0 dx \Bigl[g_{\perp}^{(v)} 
-\frac{\partial}{4\partial x}  g_{\perp}^{(a)}\Bigr](x) 
 \Bigl[-\lambda_B\int dl_- \phi_B^+(l_-)J_{p_{\gamma}}(-l_-)\\
 &&\times \Bigl(\delta_{qu} \mB_1^p (x) -\delta_{qd} \mB_2^p (x)\Bigr)+ \mB_4^p (x)\Bigr] 
+2 \delta_{qd} C_{7\gamma} \frac{\alpha}{\pi}\frac{m_bm_B}{m_{\rho}^2}  \Biggr\}. \nonumber
\end{eqnarray}
Here we include only the tree-level contributions to Cabibbo-suppressed terms with $\lambda_u^{(s)}$ in Eq.~(\ref{K2K}) because it is numerically comparable to 
the other term with $\mB_4^c$. 

In the convolutions of $\mA_1$ and $\phi_{\perp}(x)/\bar{x}$ in Eqs.~(\ref{K1K},\ref{K1rho}) and   $\mB_4$ and $g_{\perp}^{(v)}$ in Eqs.~(\ref{K2K},\ref{K2rho}),
there are endpoint divergences, which can be eliminated with the zero-bin subtractions \cite{Manohar:2006nz}   
\begin{eqnarray} 
\label{zerobin}
\int^1_0 dx \frac{\phi_{\perp} (x)}{\bar{x}^2} &\longrightarrow& \int^1_0 dx \frac{\phi_{\perp}(x) + \bar{x} \phi'_{\perp} (1)}{\bar{x}^2}, \\
\int^1_0 dx \frac{g_{\perp}^{(v)}(x)}{\bar{x}} &\longrightarrow& 
\int^1_0 dx \frac{g_{\perp}^{(v)}(x) - g_{\perp}^{(v)} (1)}{\bar{x}}. \nonumber 
\end{eqnarray}
The zero-bin subtraction removes infrared divergences from the $x$ integrals. We have dropped all finite terms including 
logarithms associated with rapidity scale dependence. We estimate the uncertainty associated with this procedure
to be 50\%.

For numerical estimates of $b_q^{K^*,\rho}$ in Eqs.~(\ref{bqK}) and (\ref{bqrho}), we use the following set of parameters: $\{m_b, m_c, m_B, m_{K^*},
m_{\rho}, f_B, f_{K^*}, f_{\rho}, f_{\perp}^{K^*}, f_{\perp}^{\rho} \}=\{4.8, 1.3, 5.28, 0.894, 0.775, 0.2\pm0.03, 0.218, 0.209, 0.175\pm0.025,
0.150\pm 0.025\}~\rm{GeV}$. The CKM parameters are $\overline{\rho} =0.221\pm0.064$ and $\overline{\eta} = 0.340\pm0.045$. All Wilson
coefficients and hard functions are evaluated at the scale $\mu=m_b$, we do not include any renormalization group evolution, and we use the
asymptotic forms for the vector meson wave function, $\phi_{\perp}$ and $g_{\perp}^{(v),(a)}$. Our estimates for the isospin asymmetries in the
absence of NPCP contributions are 
\begin{equation} 
\label{thexch} 
\Delta_{0-}^{K^*} = 0.04\pm0.02,~~~~\Delta_{0-}^{\rho} = 0.02\pm0.02,
\end{equation} 
where the dominant errors come from $\lambda_B$, $\zeta_{\perp}^V$, and CKM factors. 
These estimates for $\Delta_{0-}^V$ are comparable to previous theoretical results 
\cite{Kagan:2001zk,Beneke:2004dp,Ali:2002kw}. 
Comparing to  Eq.~(\ref{expasym}), we see that $\Delta_{0-}^{K^*}$ is consistent, but $\Delta_{0-}^{\rho}$
disagrees by about 1.7~$\sigma$.

Next we include the NPCP contribution in our calculation. In addition to the isospin asymmetries,
the NPCP can contribute to the CP violating asymmetry, $\Delta^\rho_{+-}$~\cite{Taniguchi:2008ty},
and to the branching ratio, ${\rm Br}[B^+ \to \rho^+ \gamma]$~\cite{HFAG}.  In order to obtain values 
of $m_B/\Lambda_{c\bar{c}}$ that are not inconsistent with measurements of these quantities, we perform
a least squares fit to all four observables. In our calculations of $\Delta^\rho_{+-}$ 
and ${\rm Br}[B^+\to \rho^+ \gamma]$, we include only the leading order and NPCP contributions, without
any $O(1/m_b)$ corrections.
An analysis that includes the $O(1/m_b)$ corrections to all four
observables is clearly required but is beyond the scope of this paper. 
\begin{table}[t!]
  \begin{tabular}{|c|c|c|c|}
\hline
  & \hspace{0.35 in} Exp. \hspace{0.35 in} & \hspace{0.25 in}w/o NPCP \hspace{0.25 in} & \hspace{0.25 in} w/ NPCP  \hspace{0.25 in} \\
\hline
\hline
\hspace{0.25 in}$\Delta^{K^*}_{0-}$ \hspace{0.25 in} & $0.03 \pm 0.04$ &  $0.04 \pm 0.02$&  $0.10 \pm 0.05$ \\
\hline
$\Delta^\rho_{0-}$ & $0.26 \pm 0.14$ &  $0.02 \pm 0.02$&  $0.10 \pm 0.06$ \\
\hline
$\Delta^\rho_{+-}$ & $0.11 \pm 0.33$ &  $0.08 \pm 0.02$&  $0.07 \pm 0.13$ \\
\hline
${\rm Br}[B^+\to \rho^+ \gamma] \times 10^6$ & $0.96 \pm 0.24$ &  $1.80 \pm 0.69$&  $1.63 \pm 0.67$ \\
\hline
\hline
  \end{tabular}
\caption{Theoretical predictions with and without NPCP compared to experimental data.}
\label{table}
\end{table}

The results of the fits along with experimental results are shown in Table~\ref{table}. 
The first column lists the observables considered and the second gives their measured values including
errors. In the third column, we show the theoretical prediction in the absence of the NPCP 
contribution for the values of the parameters given earlier. The last column gives the results
of the fit with the NPCP included. We extract 
\begin{eqnarray}
{\rm Re}\left[\frac{m_B}{\Lambda_{c\bar{c}}}\right] = -0.102 \pm 0.063 \, ,\qquad 
{\rm Im}\left[\frac{m_B}{\Lambda_{c\bar{c}}}\right] = 0.022 \pm 0.255 \, .
\end{eqnarray}
The $\chi^2$ of the predictions in column three of Table~\ref{table}
is 15.2. Including the NPCP, the $\chi^2$ is 12.1, so the overall agreement between experiment
and theory is slightly improved. Note that after including the NPCP the theoretical 
prediction for $\Delta^\rho_{0-}$ increases so that the 1$\sigma$ error band of the experimental
result and the theoretical result now overlap. However, the prediction for $\Delta^{K^*}_{0-}$ 
is now significantly increased. The trend suggested by the central values of the experimental data,
namely a large value of $\Delta^\rho_{0-}$ and small value of $\Delta^{K^*}_{0-}$, does not 
seem to be naturally accommodated by including NPCP contribution. However, once theoretical and experimental
uncertainties are taken into account, the theoretical predictions are consistent with both 
isospin asymmetries. For the range of values of $m_B/\Lambda_{c\bar{c}}$ obtained in our fit,
the NPCP does not have significant impact on the theoretical predictions
for $\Delta^\rho_{+-}$ and ${\rm Br}[B^+ \to \rho^+ \gamma]$. Finally, inclusion of NPCP contributions
substantially increases the uncertainty in all theoretical predictions because the parameter
$m_B/\Lambda_{c\bar{c}}$ is not well constrained.

These results indicate that the NPCP can increase the isospin violating asymmetry $\Delta^\rho_{0-}$
to bring theoretical predictions closer to current data, while maintaining consistency with the other 
observed asymmetries. It is not possible to obtain predictions for the two isospin asymmetries
that are in agreement with the central values of both $\Delta^{K^*}_{0-}$ and $\Delta^\rho_{0-}$.
 However, the uncertainties in the current measurements of all asymmetries 
are large and better measurements are need to determine whether the NPCP is an important contribution
to  $B\to \rho \gamma$ isospin and CP violating asymmetries.

To summarize, we have used SCET to calculate the isospin asymmetries in $B\to V\gamma$ decays, including all $\mO
(1/m_b)$ contributions  as well as the $\mO (v \alpha_s)$  NPCP contribution. As in nonleptonic $B$ decays
\cite{NPCP}, our analysis allows for large NPCP contributions, which could account for the  large isospin
asymmetries  measured (with large errors) in $B\to \rho\gamma$. If the isospin asymmetries are large 
and the NPCP is the source of these asymmetries, then we also expect the
NPCP to contribute to CP asymmetries in $B\to \rho \gamma$ and give larger than expected contributions to the
right-handed polarized decay rates in $B\to V\gamma$.  We speculate that the NPCP could also be responsible for the
recently measured enhanced transversely polarized decay amplitude for $B\to V V$~\cite{btovv}.  More precise
experiments will be needed to determine the exact size of NPCP and confirm these predictions.

We are thankful to Jure Zupan for initial discussions of this work. We thank Dan Pirjol and Zoltan Ligeti for useful comments on an earlier version of this paper.
C.~K.~and T.~M.~were supported in part by the Department of Energy under grant numbers DE-FG02-05ER41368 and 
DE-FG02-05ER41376.  A.~K.~L.~is supported in part by the National Science Foundation under Grant No.~PHY-0546143 and in part  by the Research Corporation.

\end{document}